\documentclass{iopconfser}

\usepackage{graphicx}
\usepackage{subcaption}
\usepackage{amsmath,amssymb}
\usepackage{booktabs}
\usepackage{hyperref} 

\begin{document}

\title{Vision Models for Medical Imaging: A Hybrid Approach for PCOS Detection from Ultrasound Scans}

\author{Md Mahmudul Hoque$^{1}$, Md Mehedi Hassain$^{2}$, Muntakimur Rahaman$^{3}$, Md. Towhidul Islam$^{4}$, Shaista Rani$^{5}$  and Md Sharif Mollah$^{6}$}

\affil{$^1$Department of CSE, CCN University of Science \& Technology, Cumilla, Bangladesh.}
\affil{$^2$Department of EEE,International Islamic University Chittagong, Chattogram, Bangladesh.}
\affil{$^3$Faculty of Engineering, Multimedia University, Cyberjaya, Malaysia.}
\affil{$^4$Department of CSE, Stamford University of Bangladesh, Dhaka, Bangladesh.}
\affil{$^5$Department of Biology, Lucknow University, Lucknow, India.}
\affil{$^6$Department of CSE, Bangladesh Army International University of Science \& Technology, Cumilla, Bangladesh.}

\email{cse.mahmud.evan@gmail.com}

\begin{abstract}
Polycystic Ovary Syndrome (PCOS) is the most familiar endocrine illness in women of reproductive age. Many Bangladeshi women suffer from PCOS disease in their older age.  The aim of our research is to identify effective vision-based medical image analysis techniques and evaluate hybrid models for the accurate detection of PCOS. We introduced two novel hybrid models combining convolutional and transformer-based approaches. The training and testing data were organized into two categories: "infected" (PCOS-positive) and "noninfected" (healthy ovaries). In the initial stage, our first hybrid model, 'DenConST' (integrating DenseNet121, Swin Transformer, and ConvNeXt), achieved 85.69\% accuracy. The final optimized model, 'DenConREST' (incorporating Swin Transformer, ConvNeXt, DenseNet121, ResNet18, and EfficientNetV2), demonstrated superior performance with 98.23\% accuracy. Among all evaluated models, DenConREST showed the best performance.  This research highlights an efficient solution for PCOS detection from ultrasound images, significantly improving diagnostic accuracy while reducing detection errors.

\end{abstract}

\section{Introduction}
Polycystic Ovary Syndrome(PCOS) \cite{shukla20} is a usual hormonal disease that effects women of reproductive age. It is marked by irregular periods, excess androgen levels, and polycistic ovaries ( enlarge ovaries with many small cyst-like folliclies).

Convolutional Neural Networks (CNNs) presented significant effectiveness in vision-based disease detection tasks \cite{nguyen2024silp, tan2025clinical, barata2021explainable}, including the identification of Polycystic Ovary Syndrome (PCOS) \cite{gollapalli2024enhancing,ghadekar2024multimodal,li2024sok}. In various studies, CNN-based models have achieved promising results, often reaching accuracies around 85\% \cite{sumathi2021study, alamoudi2023deep, karthik2024polycystic}. Notably, the ITL-CNN architecture achieved the highest reported accuracy of nearly 98\% in PCOS detection tasks \cite{gopalakrishnan2022itl}.
In \cite{hosain2022pconet}, the authors investigated PCOS, a popular endocrine disorder in women during their childbearing years that leads to hormonal imbalances and infertility. custom CNN based PCONet algorithm was developed for classifying PCOS from ovarian    ultrasound images and InceptionV3 was for fine-tuned. Evaluated on an independent test set, PCONet outperformed InceptionV3 with an accuracy of 96.56\%. The authors in \cite{kumari2021classification} developed a deep learning-based PCOS detection system using a collection of ultrasound images, with DenseNet-121, VGG-19, InceptionV3, and ResNet-50 architectures. Among these, VGG-19 achieved the highest performance 70\% accuracy in classifying PCOS from ultrasound images.

Moreover, Transformer architectures utilize self-attention mechanisms to capture global contextual relationships within the input, enabling a more comprehensive understanding of the entire image and thereby improving classification accuracy \cite{choi2023transformer}. Several studies have employed Transformer-based models for medical image analysis and disease detection, demonstrating superior performance compared to traditional approaches \cite{wang2021vit, talaatfibroidx, sri2024vit}.
The authors in \cite{pacal2004efficient} introduced an automated lung cancer detection method from CT scans utilizing deep learning, specifically focusing on the Swin Transformer architecture and a reduced window size of 6 to enhance focus on cancerous tissues, demonstrated superior performance. This model, evaluated against various CNNs and other image transformers on a public dataset and enhanced through data augmentation and transfer learning, achieved an impressive 97.58\% accuracy.

Furthermore, Recent advances in medical image analysis have illustrated the efficacy of hybrid CNN-Transformer architectures, which synergistically combine the complementary strengths of convolutional neural networks (CNNs) and Transformer models. This integration has garnered significant research interest, as CNNs excel at local feature extraction while Transformers capture long-range dependencies, offering superior performance in complex medical imaging tasks \cite{qezelbash2025hybrid}. Particularly in polycystic ovary syndrome (PCOS) detection and analysis, hybrid approaches have shown promising results. Several studies have successfully implemented these architectures, demonstrating their potential to improve diagnostic accuracy and feature representation \cite{qezelbash2025hybrid,kumar2024hybrid,li2024multi}. The combination of CNN's hierarchical feature learning with Transformer's global attention mechanisms has proven particularly effective for analyzing ovarian ultrasound images, where both local morphological details and global structural relationships are clinically significant.
In \cite{qezelbash2025hybrid}, the authors proposed ConvTransGFusion, a novel hybrid model that integrates ConvNeXt's local feature extraction with Swin Transformer an adaptive dual-attention fusion mechanism, achieving 98.9\% accuracy. The model was evaluated on six medical imaging datasets, including X-rays, ultrasounds, and MRIs.

Traditional machine learning approaches have demonstrated notable success in disease detection tasks, including the identification of Polycystic Ovary Syndrome (PCOS) and other medical conditions \cite{ahmed2023review, denny2019hope, hdaib2022detection, thakre2020pcocare}. Among these methods, Random Forest and Xgboost has consistently emerged as one of the most effective classifiers in PCOS-related studies, owing to its robustness and high predictive accuracy\cite{suha2022extended}.
The authors in \cite{hassan2020comparative} proposed PCOS detection from clinical patient data with SVM, CART, naive Bayes classification, random forest, and logistic regression. Among these, the random forest classifier achieved the highest accuracy of 96\%.

\section{Methodology}
\begin{figure}
    \centering
    \includegraphics[width=0.6\linewidth]{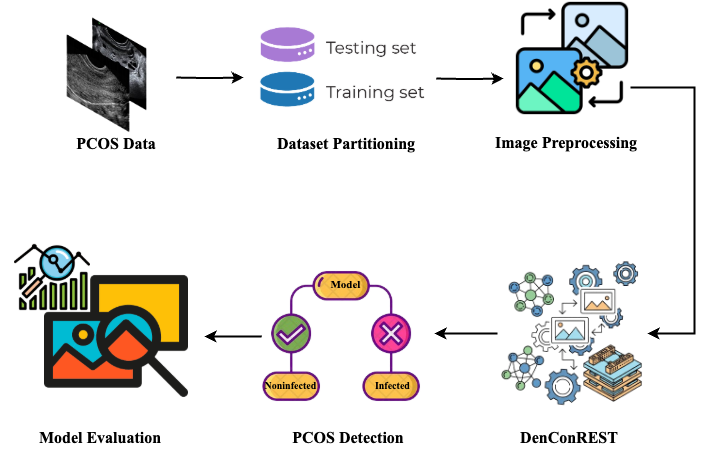}
    \caption{System Architecture}
    \label{system}
\end{figure}

Fig \ref{system} shows the proposed framework for PCOS detection from ultrasound scan images.

\subsection{Data Collection}
\begin{figure}[htbp]
    \centering
    \begin{subfigure}[b]{0.3\linewidth}
        \centering
        \includegraphics[width=\linewidth]{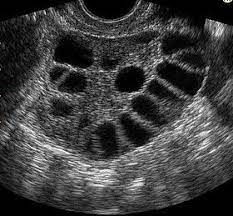}
        \caption{Infected Ultrasound Image}
        \label{fig:img1}
    \end{subfigure}
    
    \vspace{0.3cm} 
    
    \begin{subfigure}[b]{0.3\linewidth}
        \centering
        \includegraphics[width=\linewidth]{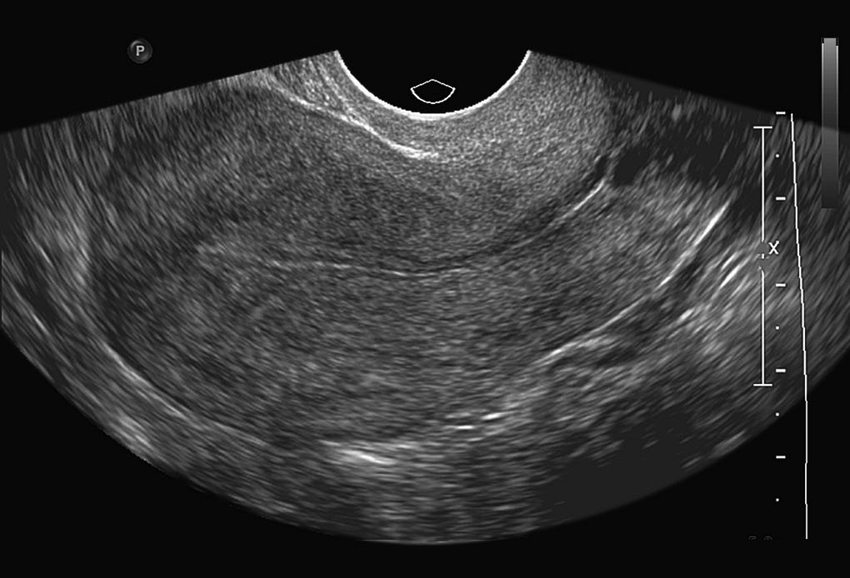}
        \caption{Noninfected Ultrasound Image}
        \label{fig:img2}
    \end{subfigure}
     
    \caption{PCOS infected and noninfected ultrasound images}
    \label{fig:combined}
\end{figure}

The PCOS ultrasound image dataset was collected from Kaggle \cite{kagglePCOSDetection}. The data folder consists of 'train' and 'test' subfolders, each containing two categories: 'infected' and 'notinfected'. The infected folder contains images of ovaries with PCOS, while the notinfected folder contains images of healthy ovaries. In the train directory, the infected folder contains 781 images, and the notinfected folder contains 1,143 images. The test directory includes 787 infected images and 1,145 noninfected images. Figure \ref{fig:combined} illustrates ultrasound scans of both infected and noninfected ovaries. Visual differences between infected and noninfected ovaries are apparent. In Figure \ref{fig:img1}, the ultrasound image shows characteristic holes (follicles), which distinguish it from Figure \ref{fig:img2} (noninfected), where no such holes are present.

\subsection{Image Preprocessing}
To establish a robust foundation for training a deep neural network to detect Polycystic Ovary Syndrome (PCOS) from ultrasound images, we implemented a preprocessing step to filter out corrupted or invalid images. This step is crucial because corrupted images can lead to errors during model training or negatively impact performance. First we have used Python Pillow (PIL) library to scan recursively to open and verify the image integrity \cite{clark2015pillow}. It scanned through all subdirectories and removed corrupted or unreadable images. This preprocessing step improves dataset reliability by removing problematic images before they can affect model training. Since medical imaging data, such as ultrasounds, can sometimes be incomplete or corrupted during acquisition or storage, this automated filtering process helps maintain data quality and consistency, leading to more robust and accurate deep learning models for PCOS detection. For maintaining aspect ratio and Ensuring uniform input size for the neural network all input images resized to a fixed dimension of 224x224 pixels. The images are then converted into PyTorch tensors \cite{mishra2022introduction}, transforming them from PIL image format (height×width×channels) to the required tensor format (channels×height×width) while simultaneously scaling pixel values to a 0-1 range. The final step involves normalization using ImageNet statistics , the mean was 0.485, 0.456, 0.406 and the std was 0.229, 0.224, 0.225. The standardization process centers the data around zero and scales it to unit variance and the comprehensive preprocessing pipeline ensures that the ultrasound images are optimally prepared for feature extraction and classification.

\subsection{Model Architecture}
This paper presents a deep learning framework for PCOS detection from ultrasound images, implementing both convolutional and transformer-based architectures. Five pretrained models such as EfficientNetV2, ResNet18, DenseNet121, Swin Transformer, and ConvNeXt were fine-tuned through transfer learning, final layers adapted for binary classification, PCOS or Normal\cite{woo2023convnext, yu2019abnormality, swaminathan2021multiple, tan2021efficientnetv2, liu2021swin}. The system employs standardized preprocessing (resizing, normalization) and optimized training with Adam (lr=0.0001-0.001) and CrossEntropyLoss. To enhance the robust model performance, we proposed DenConST and DenConREST, where DenConST combines predictions from ConvNeXt, Swin Transformer and DenseNet121, on the other hand DenConREST combines predictions from all architectures through averaging, capitalizing on their complementary strengths. Comprehensive evaluation metrics (accuracy, precision, recall, F1) demonstrate the comparative performance of individual models and the ensemble. This multi-architecture approach ensures high diagnostic accuracy while providing insights into model-specific strengths for medical image analysis. The implementation uses PyTorch and TIMM libraries , ensuring reproducibility and scalability for clinical applications.

\subsection{Training Configuration and Optimization}
The training process employed the Adam optimizer with learning rates carefully selected for each architecture (ranging from 0.0001 to 0.001) to ensure stable convergence. We used CrossEntropyLoss as our objective function, which is standard for multi-category classification \cite{zhang2021competing}. Batch size was fixed at 32 across all experiments to maintain consistency while balancing memory constraints and gradient estimation quality. he models were trained for 100 epochs, with the exact number adjusted based on model complexity and observed convergence behavior. During training, we implemented a standard forward-backward propagation loop: computing predictions, calculating loss, zeroing gradients, performing backpropagation, and updating weights through the optimizer. This process was repeated for all batches in each epoch, with training loss monitored to track learning progress.

\subsection{Hybrid Model Development}
\begin{figure}
    \centering
    \includegraphics[width=0.6\linewidth]{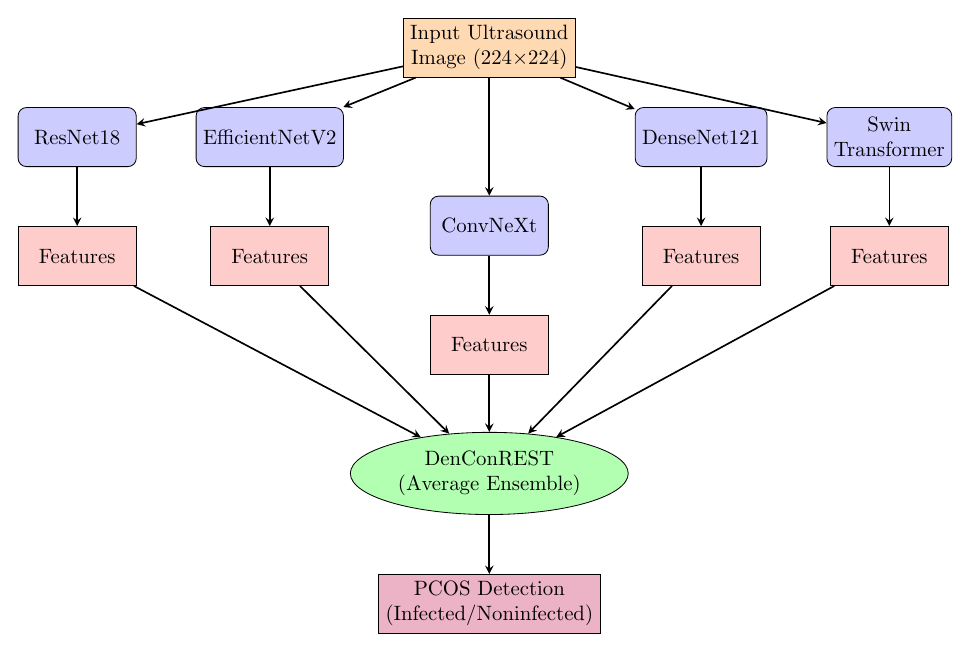}
    \caption{DenConREST Architecture}
    \label{model}
\end{figure}

To combine the strengths of multiple architectures, we developed DenConST and DenConREST, two advanced Hybrid approach. DenConST is a custom PyTorch module integrates predictions from ConvNeXt, Swin Transformer and DenseNet121 and DenConREST is a custom PyTorch module integrates predictions from all five individual models (Swin Transformer, ConvNeXt, DenseNet121, ResNet18, and EfficientNetV2) through a simple averaging mechanism. During forward passes, each constituent model processes the input image independently, and their output logits are averaged element-wise. This hybrid  strategy capitalizes on the diverse feature extraction capabilities of different architectures - from CNN-based local feature detection in ResNet to global attention mechanisms in Swin Transformer - potentially yielding more robust predictions than any single model alone \cite{qezelbash2025hybrid}. The hybrid  was designed to be computationally efficient during inference by running models in parallel where hardware permits. Fig \ref{model} illustrates the hybrid  model architecture.

Here is the Mathematical formulation for the hybrid  model:

Let:
\begin{itemize}
    \item $x$ = input ultrasound image
    \item $f_i(x)$ = prediction logits of model $i$
    \item $w_i$ = learned weight for model $i$
\end{itemize}

The hybrid  output $F(x)$ is:

\begin{equation}
F(x) = \sigma\left(\sum_{i=1}^5 w_i \cdot f_i(x)\right)
\end{equation}

Where:
\begin{equation*}
\begin{cases}
f_1(x) & = \text{Swin Transformer (ST)} \\
f_2(x) & = \text{ConvNeXt (Con)} \\
f_3(x) & = \text{DenseNet121 (Den)} \\
f_4(x) & = \text{ResNet18 (R)} \\
f_5(x) & = \text{EfficientNetV2 (E)} \\
\sum w_i & = 1 \quad \text{(weight normalization)} \\
\sigma & = \text{Sigmoid activation}
\end{cases}
\end{equation*}

\subsection{Evaluation Methodology}
We conducted comprehensive evaluations using four performance indicators: accuracy, precision, recall, and F1 score. Each model was evaluated on the same holdout dataset to ensure fair comparison. The evaluation process involved passing test images through each model without gradient computation, collecting predicted class probabilities, and comparing these against ground truth labels. For the hybrid, we implemented a forward pass that aggregated predictions from all component models before computing metrics. This rigorous evaluation framework allows us to assess not just overall accuracy but also clinically relevant aspects like false positive rates (precision) and sensitivity (recall), which are crucial for medical diagnosis systems.

\subsection{Implementation Details}
The experiments were employed on Kaggle’s cloud platform using an NVIDIA Tesla P100 GPU (PCIe) with 16GB of HBM2 memory and 3,584 CUDA cores, providing 9.3 TFLOPS of FP32 performance. The software stack included PyTorch 1.12.0 and TIMM  0.6.7 with CUDA 11.2 and cuDNN 8.1.0 for accelerated deep learning operations. To optimize performance within Kaggle’s computational constraints, we employed mixed-precision training (AMP) to reduce memory overhead while maintaining numerical stability, allowing efficient training of memory-intensive models like Swin Transformer and ConvNeXt. Batch sizes were set to 32 for CNNs (ResNet, DenseNet, EfficientNetV2) and 16 for transformer-based models to avoid out-of-memory errors, with gradient accumulation used where necessary. The DataLoader was configured with 2 CPU workers (Kaggle’s maximum allowance) and pinned memory to minimize data transfer bottlenecks. Periodic torch.cuda.empty\_cache() calls prevented memory leaks, and model checkpoints were saved to Kaggle’s persistent storage to accommodate the 12-hour session limit. This setup ensured efficient training and evaluation of all models, including the ensemble system, while maintaining reproducibility through fixed random seeds and containerized dependencies. The P100’s high memory bandwidth (732 GB/s) proved particularly beneficial for vision transformers, enabling stable training despite their large parameter counts.

\section{Results Analysis}

The proposed model for Polycystic Ovary Syndrome (PCOS) detection utilizes a hybrid approach combining convolutional and transformer-based architectures. Five pretrained models - EfficientNetV2, ResNet18, DenseNet121, Swin Transformer, and ConvNeXt - were employed for binary classification of ultrasound images as either infected or noninfected. We developed two hybrid architectures: DenConST combines DenseNet121, Swin Transformer, and ConvNeXt, while DenConREST integrates all five models. Performance was evaluated using accuracy, precision, recall, and F1-score metrics on a test set of 1,922 preprocessed images (781 noninfected and 1,141 infected). All models achieved over 50\% accuracy, with EfficientNetV2 showing the strongest performance at 79\% accuracy, 92\% precision, and 80\% F1-score. Notably, ResNet18 achieved 100\% recall, the highest among all individual models.

\begin{table}[h]
\caption{Performance Metrics of PCOS Detection Models}
\centering
\begin{tabular}{@{}lcccc@{}}\toprule
Model & Accuracy & Precision & Recall & F1 Score \\\midrule
Swin Transformer & 0.5645 & 0.5896 & 0.8764 & 0.7050 \\
ConvNeXt & 0.5884 & 0.5994 & 0.9246 & 0.7273 \\
DenseNet121 & 0.6883 & 0.6879 & 0.8694 & 0.7681 \\
ResNet18 & 0.5937 & 0.5937 & 1.0000 & 0.7450 \\
EfficientNetV2 & 0.7955 & 0.9279 & 0.7108 & 0.8050 \\
DenConST (Hybrid) & 0.8569 & 0.8832 & 0.8747 & 0.8789 \\
DenConREST (Hybrid) & \textbf{0.9823} & \textbf{0.9719} & \textbf{0.9991} & \textbf{0.9849} \\
\bottomrule
\end{tabular}
\label{tab:pcos_metrics}
\end{table}

Nevertheless, Our proposed DenConREST demonstrated superior performance with 98.2\% accuracy, 97.1\% precision, 99.9\% recall, and 98.4\% F1-score, as detailed in Table \ref{tab:pcos_metrics}. The secondary model, DenConST, achieved the second-best results with 85.6\% accuracy and 87.8\% F1-score, along with 88.3\% precision and 87.4\% recall. These results highlight the proficiency of combining multiple architectures for improved PCOS detection.

\begin{figure}
    \centering
    \includegraphics[width=0.6\linewidth]{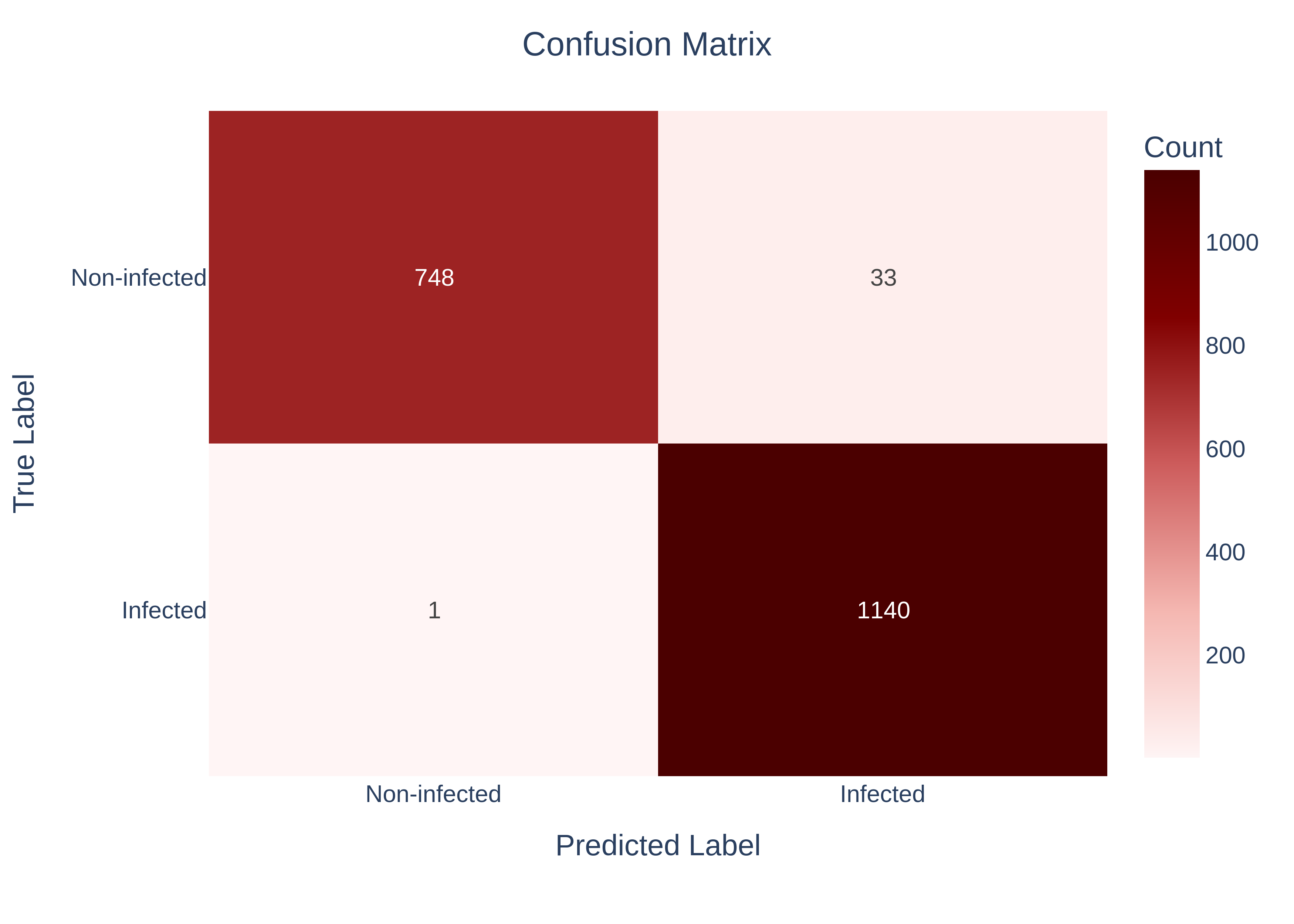}
    \caption{Confusion Matrix of DenConREST Model}
    \label{Conf}
\end{figure}

Figure \ref{Conf} illustrates the confusion matrix of the DenConREST model for PCOS detection from ultrasound images. The confusion matrix was plotted using the test dataset and DenConREST. Among 1141 infected images, our proposed model accurately classified 1140 images and only 1 image couldn't be classified accurately, which made the model the best and achieved 99.9\% recall. Moreover, among 781 noninfected images, our proposed model classified 748 images accurately and only 33 images were misclassified as infected.This performance demonstrates the model's robustness in both architectural design and clinical application for PCOS detection through ultrasound imaging.

\section{Discussion}
Compared to earlier works that relied primarily on single CNN architectures with accuracies around 85–90\% , our hybrid architecture achieves superior performance. In a single model, whenever accuracy is found to be high, it may not work properly in real scenario. However, a hybrid model works better than a single model whenever in complex operation. This suggests its suitability for clinical applications where high sensitivity is crucial to avoid false negatives in PCOS diagnosis. DenConREST maintains balance across all metrics, minimizing both types of classification errors. The experimental results demonstrate that the proposed DenConREST model significantly outperforms all cnn and hybrid models across  all evaluation metrics. With an accuracy of 98.2\%, DenConREST establishes a new benchmark for PCOS detection using ultrasound images. Notably, the hybrid approach benefits from DenseNet121's efficient feature propagation, ConvNeXt’s depthwise convolutional refinement, and Swin Transformer’s hierarchical attention, leading to a more discriminative and generalizable model. Clinically, the ability to accurately detect PCOS from ultrasound images with minimal human intervention could assist radiologists in faster screening, especially in limited resource.

\section{Conclusion}
This research introduces DenConREST, a hybrid  model that combines convolutional neural networks (CNNs) and transformer-based architectures for automated PCOS detection from ultrasound images. This work integrates the predictive capabilities of EfficientNetV2, ResNet18, DenseNet121, Swin Transformer, and ConvNeXt using an optimized weighted ensemble technique to enhance overall classification performance. The hybrid model achieved 98.23\% accuracy, 99.9\% recall, and a 98.4\% F1-score, demonstrating that carefully designed hybrid architectures can significantly improve automated ultrasound analysis and offer clinicians a reliable decision-support tool. The model's high recall ensures that nearly all PCOS cases are correctly identified, which is critical for timely diagnosis and intervention. To further enhance clinical applicability, future work will focus on multicenter validation, expanding the dataset with multi-center studies, and incorporating hormonal biomarkers for multimodal diagnosis. The proposed framework contributes to the advancement of automated PCOS diagnosis by effectively addressing the shortage of imaging specialists and meeting the growing diagnostic demands in under-resourced healthcare facilities.




\begin{thebibliography}{14}
\bibitem{ahmed2023review} Ahmed, S., Rahman, M.S., Jahan, I., et al.: 'A review on the detection techniques of polycystic ovary syndrome using machine learning', IEEE Access, 2023, 11, pp. 86522--86543

\bibitem{suha2022extended}
Suha, S. A. and Islam, M. N.: `An extended machine learning technique for polycystic ovary syndrome detection using ovary ultrasound image', \textit{Scientific Reports}, 2022, \textbf{12}(1), pp. 17123
\bibitem{clark2015pillow} A. Clark et al., “Pillow (pil fork) documentation,” readthedocs, 2015.

\bibitem{kumar2024hybrid} N. A. R. Kumar and V. Varadarajan, “Hybrid pcos net: A synergistic
cnn-lstm approach for accurate polycysti covary syndrome detection,”
2024.

\bibitem{yu2019abnormality} X. Yu and S.-H. Wang, “Abnormality diagnosis in mammograms by
transfer learning based on resnet18,” Fundamenta Informaticae, vol. 168, no. 2-4, pp. 219–230, 2019.

\bibitem{swaminathan2021multiple} A. Swaminathan, C. Varun, S. Kalaivani et al., “Multiple plant leaf disease classification using densenet-121 architecture,” Int. J. Electr. Eng. Technol, vol. 12, no. 5, pp. 38–57, 2021.


\bibitem{tan2021efficientnetv2} M. Tan and Q. Le, “Efficientnetv2: Smaller models and faster training,” in International conference on machine learning. PMLR, 2021, pp.10 096–10 106.


\bibitem{liu2021swin} Z. Liu, Y. Lin, Y. Cao, H. Hu, Y. Wei, Z. Zhang, S. Lin, and B. Guo, “Swin transformer: Hierarchical vision transformer using shifted windows,” in Proceedings of the IEEE/CVF international conference on
computer vision, 2021, pp. 10 012–10 022.





\bibitem{li2024sok}
Li, M., He, Z., Nie, L., Shi, L., Lin, M., Li, M., Cheng, Y., Liu, H. and Xue, L.: `SoK: Intelligent Detection for Polycystic Ovary Syndrome (PCOS)', \textit{medRxiv}, 2024, pp. 2024--12, Cold Spring Harbor Laboratory Press


\bibitem{thakre2020pcocare}
Thakre, V., Vedpathak, S., Thakre, K., and Sonawani, S.: `PCOcare: PCOS detection and prediction using machine learning algorithms', *Biosci. Biotechnol. Res. Commun.*, 2020, **13**(14), pp. 240--244


\bibitem{li2024multi}
Li, Y., Zhao, B., Wen, L., Huang, R., and Ni, D.: `Multi-purposed diagnostic system for ovarian endometrioma using CNN and transformer networks in ultrasound', *Biomed. Signal Process. Control*, 2024, **91**, pp. 105923


\bibitem{alamoudi2023deep} Alamoudi, A., Khan, I.U., Aslam, N., et al.: 'A deep learning fusion approach to diagnose polycystic ovary syndrome (PCOS)', Appl. Comput. Intell. Soft Comput., 2023, 2023, (1), pp. 9686697

\bibitem{barata2021explainable} Barata, C., Celebi, M.E., Marques, J.S.: 'Explainable skin lesion diagnosis using taxonomies', Pattern Recognit., 2021, 110, pp. 107413

\bibitem{gopalakrishnan2022itl} Gopalakrishnan, C., Iyapparaja, M.: 'ITL-CNN: Integrated transfer learning-based convolution neural network for ultrasound PCOS image classification', Int. J. Pattern Recognit. Artif. Intell., 2022, 36, (16), pp. 2240002

\bibitem{hassan2020comparative} Hassan, M.M., Mirza, T.: 'Comparative analysis of machine learning algorithms in diagnosis of polycystic ovarian syndrome', Int. J. Comput. Appl., 2020, 975, (8887)

\bibitem{li2024multi} Li, Y., Zhao, B., Wen, L., et al.: 'Multi-purposed diagnostic system for ovarian endometrioma using CNN and transformer networks in ultrasound', Biomed. Signal Process. Control, 2024, 91, pp. 105923

\bibitem{nguyen2024silp} Nguyen, K.D., Zhou, Y.H., Nguyen, Q.V., et al.: 'SILP: enhancing skin lesion classification with spatial interaction and local perception', Expert Syst. Appl., 2024, 258, pp. 125094

\bibitem{tan2025clinical} Tan, L., Wu, H., Zhu, J., et al.: 'Clinical-inspired skin lesions recognition based on deep hair removal with multi-level feature fusion', Pattern Recognit., 2025, 161, pp. 111325

\bibitem{wang2021vit} Wang, H., Ji, Y., Song, K., et al.: 'ViT-P: Classification of genitourinary syndrome of menopause from OCT images based on vision transformer models', IEEE Trans. Instrum. Meas., 2021, 70, pp. 1--14

\bibitem{zhang2021competing} Zhang, K., Guo, Y., Wang, X., et al.: 'Competing ratio loss for discriminative multi-class image classification', Neurocomputing, 2021, 464, pp. 473--484



\bibitem{hdaib2022detection}
Hdaib, D., Almajali, N., Alquran, H., Mustafa, W. A., Al-Azzawi, W., and Alkhayyat, A.: `Detection of polycystic ovary syndrome (PCOS) using machine learning algorithms', *Proc. 5th Int. Conf. on Engineering Technology and its Applications (IICETA)*, 2022, pp. 532--536

\bibitem{sri2024vit}
Sri, P.L., Madhumita, M.R., Madhuraj, T., Goel, A., Aadhithya, A., and Soman, K.P.: `ViT-PCOS: Vision Transformer for Automated Polycystic Ovary Syndrome Detection'. Proc. 4th Int. Conf. Emerging Frontiers in Electrical and Electronic Technologies (ICEFEET), 2024, pp. 1--6

\bibitem{denny2019hope} Denny, A., Raj, A., Ashok, A., et al.: 'i-HOPE: Detection and prediction system for polycystic ovary syndrome (PCOS) using machine learning techniques'. Proc. TENCON 2019, IEEE Reg. 10 Conf., 2019, pp. 673--678

\bibitem{gollapalli2024enhancing} Gollapalli, S., Kanchanamala, P., Muppidi, S., et al.: 'Enhancing the Detection of PCOS using Deep Learning'. Proc. ICSSAS, 2024, pp. 736--740

\bibitem{ghadekar2024multimodal} Ghadekar, P., Tekade, S., Sakharwade, D., et al.: 'Multimodal PCOS Detection: Combining XGBoost for Images with Zero Shot Learning for Textual Data'. Proc. APCIT, 2024, pp. 1--8

\bibitem{hosain2022pconet} Hosain, A.K.M.S., Mehedi, M.H.K., Kabir, I.E.: 'PCONet: A convolutional neural network architecture to detect PCOS from ovarian ultrasound images'. Proc. Int. Conf. Eng. Emerg. Technol. (ICEET), 2022, pp. 1--6

\bibitem{karthik2024polycystic} Karthik, Y., Sruthi, R., Sujithra, M.: 'Polycystic Ovary Syndrome Prediction through CNN based Image Analysis: A Deep Learning Based Approach'. Proc. I-SMAC, 2024, pp. 1547--1553

\bibitem{liu2021swin} Liu, Z., Lin, Y., Cao, Y., et al.: 'Swin Transformer: Hierarchical Vision Transformer using Shifted Windows'. Proc. IEEE/CVF Int. Conf. Comput. Vis., 2021, pp. 10012--10022

\bibitem{mao2023cross} Mao, A., Mohri, M., Zhong, Y.: 'Cross-entropy loss functions: Theoretical analysis and applications'. Proc. ICML, 2023, pp. 23803--23828



\bibitem{sumathi2021study} Sumathi, M., Chitra, P., Prabha, R.S., et al.: 'Study and detection of PCOS related diseases using CNN'. Proc. IOP Conf. Ser. Mater. Sci. Eng., 2021, 1070, (1), pp. 012062

\bibitem{woo2023convnext} Woo, S., Debnath, S., Hu, R., et al.: 'ConvNeXt V2: Co-designing and scaling convnets with masked autoencoders'. Proc. CVPR, 2023, pp. 16133--16142


\bibitem{mishra2022introduction} Mishra, P.: 'Introduction to PyTorch, Tensors, and Tensor Operations', in 'PyTorch Recipes: A Problem-Solution Approach to Build, Train and Deploy Neural Network Models' (Springer, 2022), pp. 1--28

\bibitem{shukla20} Shukla A, Rasquin LI, Anastasopoulou C. Polycystic Ovarian Syndrome. [Updated 2025 May 4]. In: StatPearls [Internet]. Treasure Island (FL): StatPearls Publishing; 2025 Jan-. Available from: https://www.ncbi.nlm.nih.gov/books/NBK459251/

\bibitem{choi2023transformer}
Choi, S.R., and Lee, M.: `Transformer architecture and attention mechanisms in genome data analysis: a comprehensive review', *Biology*, 2023, **12**, (7), pp. 1033

\bibitem{qezelbash2025hybrid}
Qezelbash-Chamak, J., and Hicklin, K.: `A Hybrid Learnable Fusion of ConvNeXt and Swin Transformer for Optimized Image Classification', *IoT*, 2025, **6**, (2), pp. 30




\bibitem{kumari2021classification} Kumari, S.: 'Classification of PCOS/PCOD using Transfer Learning and GAN Architectures to Generate Pseudo Ultrasound Images'. PhD thesis, National College of Ireland, 2021

\bibitem{pacal2004efficient}
Pacal, I., and Universitesi, I.: `Efficient Swin Transformer Model for Accurate Detection of Lung Cancer'. ResearchGate, 2004

\bibitem{talaatfibroidx}
Talaat, F. M.: `FibroidX: Vision Transformer-Powered Prognosis and Recurrence Prediction for Uterine Fibroids', \textit{(ResearchGate)}, 2024.



\bibitem{kagglePCOSDetection} `PCOS detection using ultrasound images', \url{https://www.kaggle.com/datasets/anaghachoudhari/pcos-detection-using-ultrasound-images}, accessed 03-06-2025

\end{thebibliography}
\end{document}